\documentclass{iau-JDSS}
\usepackage[pdftex]{graphicx}

\title[Multiplicity of Exoplanet Host Stars] 
{Multiplicity Study of Exoplanet Host Stars}

\author[M. Mugrauer et al.]   
{M. Mugrauer, R. Neuh\"{a}user, C. Ginski, T. Eisenbeiss}

\affiliation{Astrophysical Institute and University-Observatory Jena, Germany \break email: markus@astro.uni-jena.de\\[\affilskip]}

\pubyear{2009}
\volume{Volume 15}  
\pagerange{1-2}
\jname{Highlights of Astronomy, Volume 15}
\editors{Ian F Corbett, ed.}
\begin{document}

\maketitle

\begin{abstract}
We present recent results of our ongoing multiplicity study of exoplanet host stars.

\keywords{exoplanets, multiple stars, planet formation}
\end{abstract}

\firstsection 

\section{New low-mass stellar companions of exoplanet host stars}

In our imaging campaign, carried out with SofI/NTT and UFTI/UKIRT, we directly detected so far
several new companions of exoplanet host stars. Among them HD\,3651\,B the first T dwarf companion
of an exoplanet host star (\cite{1} \& \cite{2}), HD\,27442\,B a white dwarf which is the secondary
of the most evolved exoplanet host star system presently known (\cite{3}), as well as the binary
companion of HD\,65216, whose B component is a low-mass star, while HD\,65216\,C is either a
massive brown dwarf or a very low-mass star (\cite{4}).

Recently, we identified two new low-mass stellar companions of the exoplanet host stars HD\,125612
and HD\,212301. The co-moving companion of HD\,125612 is a wide M4 dwarf (0.18\,M$_{\odot}$),
located about 4750\,AU southeast of its primary. The co-moving companion of HD\,212301 is a close
M3 dwarf (0.35\,M$_{\odot}$), which was found by us about 230\,AU north-west of the exoplanet host
star. The binaries HD\,125612\,AB and HD\,212301\,AB are two new members in the continuously
growing list of exoplanet host star systems of which more than 40 are presently known (\cite{5}).

\section{Lucky-Imaging search for close companions of exoplanet host stars}

We started a search for close stellar companions of exoplanet host stars at the Calar Alto
Observatory in Spain, using the Lucky-Imaging technique. The observations are carried out with the
2.2\,m telescope and its Lucky-Imaging camera AstraLux in the I-band. We always take several
thousand short integrated images with integration times down to 30\,ms, and choose a total
integration time of about 30\,min per target. After the standard data-reduction, our Lucky-Imaging
pipeline measures the Strehl-ratios of all images, and then selects only those images with the
highest Strehl-ratios (selection rates from 1 to 10\%). Finally, all selected images are shifted
and combined. According to the achieved AstraLux detection limit, beyond 1\,arcsec ($\sim$40\,AU of
projected separation at the average distance of our targets) we are sensitive for all stellar
companions ($>0.08$\,M$_{\odot}$) around our targets. Hence, close stellar companions, which remain
invisible in seeing limited observations, are clearly detectable in our AstraLux imaging campaign.

\begin{appendix}

\section{New companions of exoplanet host stars}

\begin{figure}[h!]
\resizebox{\hsize}{!}{\includegraphics[]{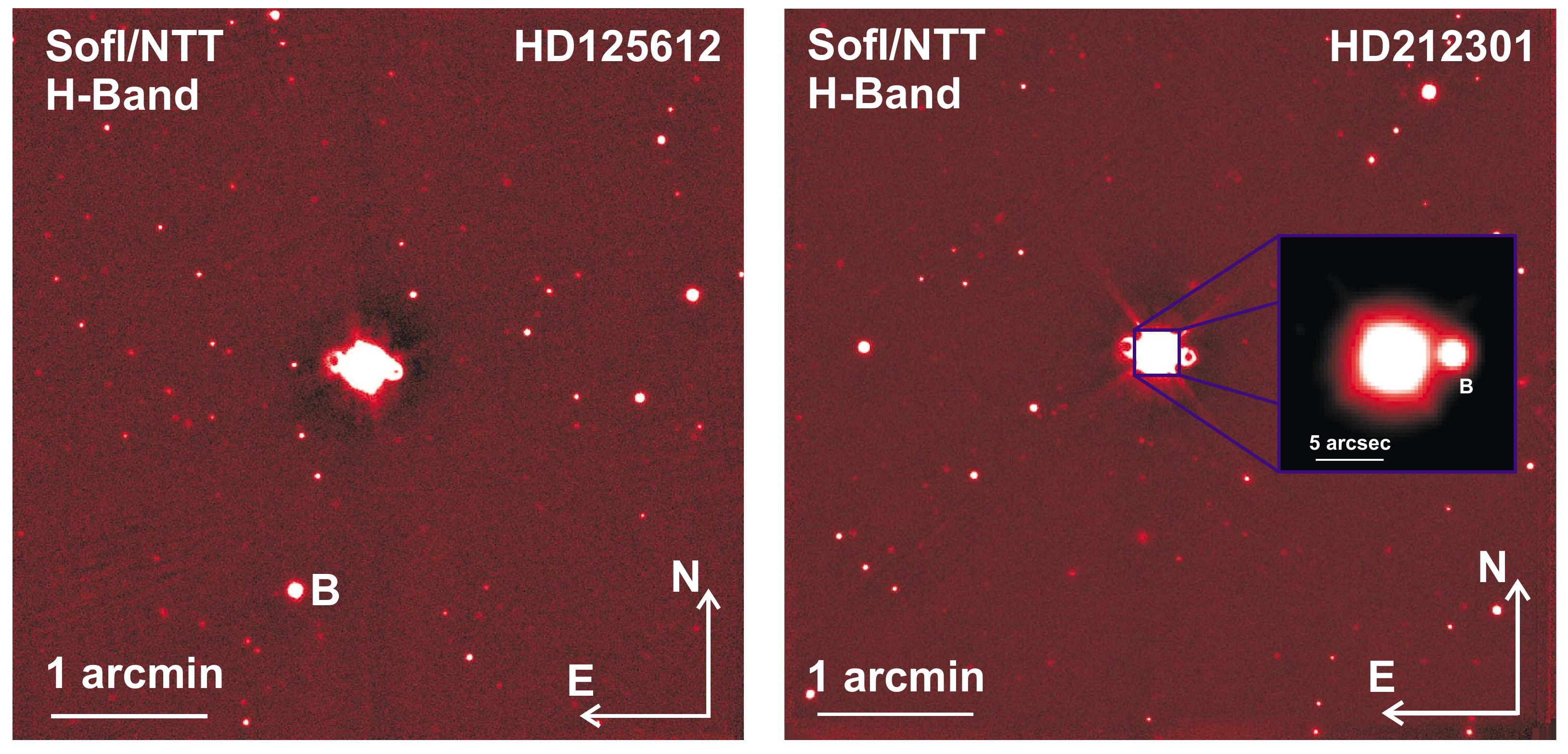}}
\resizebox{\hsize}{!}{\includegraphics[]{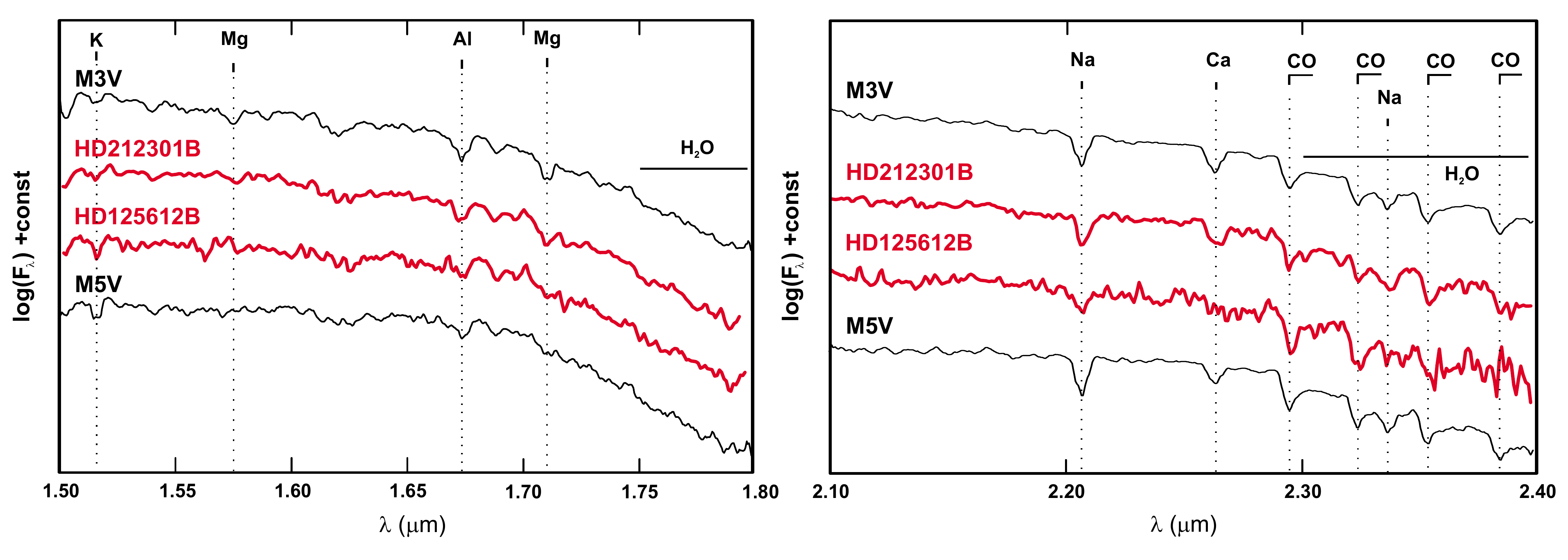}} \caption{\textbf{Top:} In the course of our
SofI/NTT imaging campaign, recently we identified two new low-mass stellar companions of the
exoplanet host stars HD\,125612 and HD\,212301. Both companions clearly share the proper motions of
their primaries, derived with the comparison of our SofI images, taken at different observing
epochs, or by comparing our SofI data with 2MASS images, taken several years before our SofI
observations. HD\,125612\,B is located about 1.5\,arcmin south-east of its primary. HD\,212301\,B
is located 4.4\,arcsec north-west of the planet host star. \textbf{Bottom:} SofI/NTT H- and K-band
follow-up spectra of HD\,125612\,B and HD\,212301\,B.}
\end{figure}

\begin{figure}[h!]
\resizebox{\hsize}{!}{\includegraphics[]{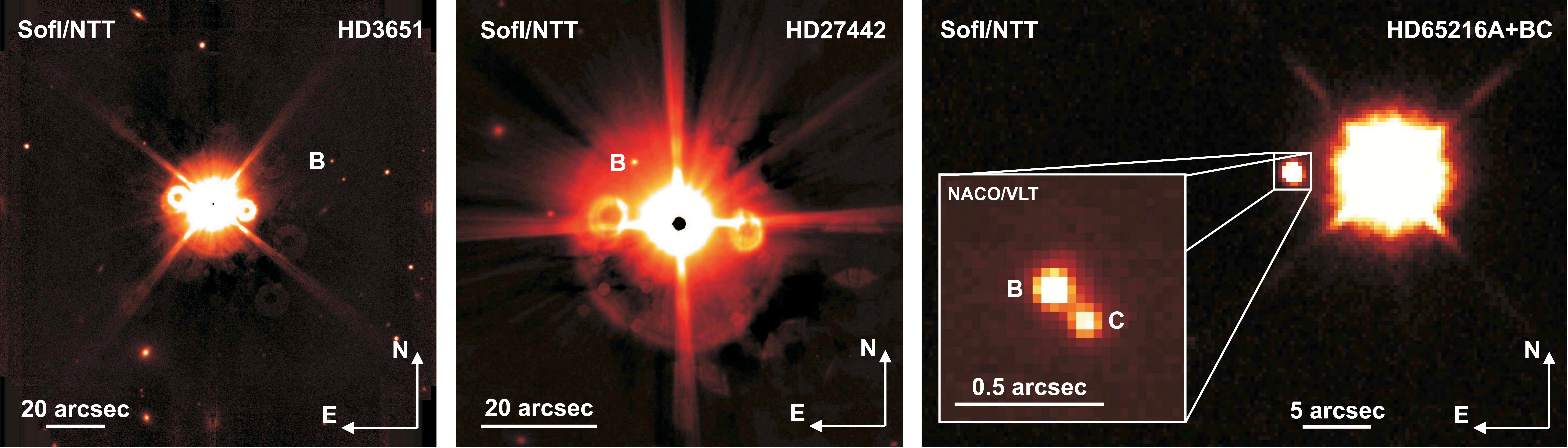}} \caption{\textbf{Left:} The faint T-dwarf
companion (SpT T7-8) to the planet host star HD3651, detected by us with SofI/NTT and UFTI/UKIRT in
the H-band. The companion clearly shares the proper motion of its primary from which it is
separated by about 43\,arcsec ($\sim$480\,AU of projected separation). \textbf{Middle:} The
co-moving companion of the exoplanet host star HD\,27442, imaged with SofI/NTT in the H-band.
HD\,27442\,B is located 12.9\,arcsec ($\sim$240\,AU of projected separation) north-east of the
planet host star. The photometry of the companion is fully consistent with a relatively young white
dwarf ($T \sim 14400$\,K, and cooling age of about 220\,Myr), confirmed by spectroscopy.
HD\,27442\,B shows Hydrogen absorption features of the Balmer, Paschen, and Bracket series in its
optical and infrared spectra. With its subgiant primary and the white dwarf companion,
HD\,27442\,AB is the most evolved exoplanet host star system presently known. \textbf{Right:} The
exoplanet host star HD\,65216 with its binary companion HD\,65216\,BC, located about 7\,arcsec
($\sim$250\,AU of projected separation) east of its primary. While HD\,65216\,B
(0.089\,M$_{\odot}$, SpT M7-8) is a low-mass star, HD\,65216\,C (0.078\,M$_{\odot}$, SpT L2-3)
could be either a massive brown dwarf or a very low-mass star.}
\end{figure}

\newpage

\section{Lucky-Imaging search for close stellar companions of exoplanet host stars with AstraLux}

\begin{figure}[h!]
\resizebox{\hsize}{!}{\includegraphics[]{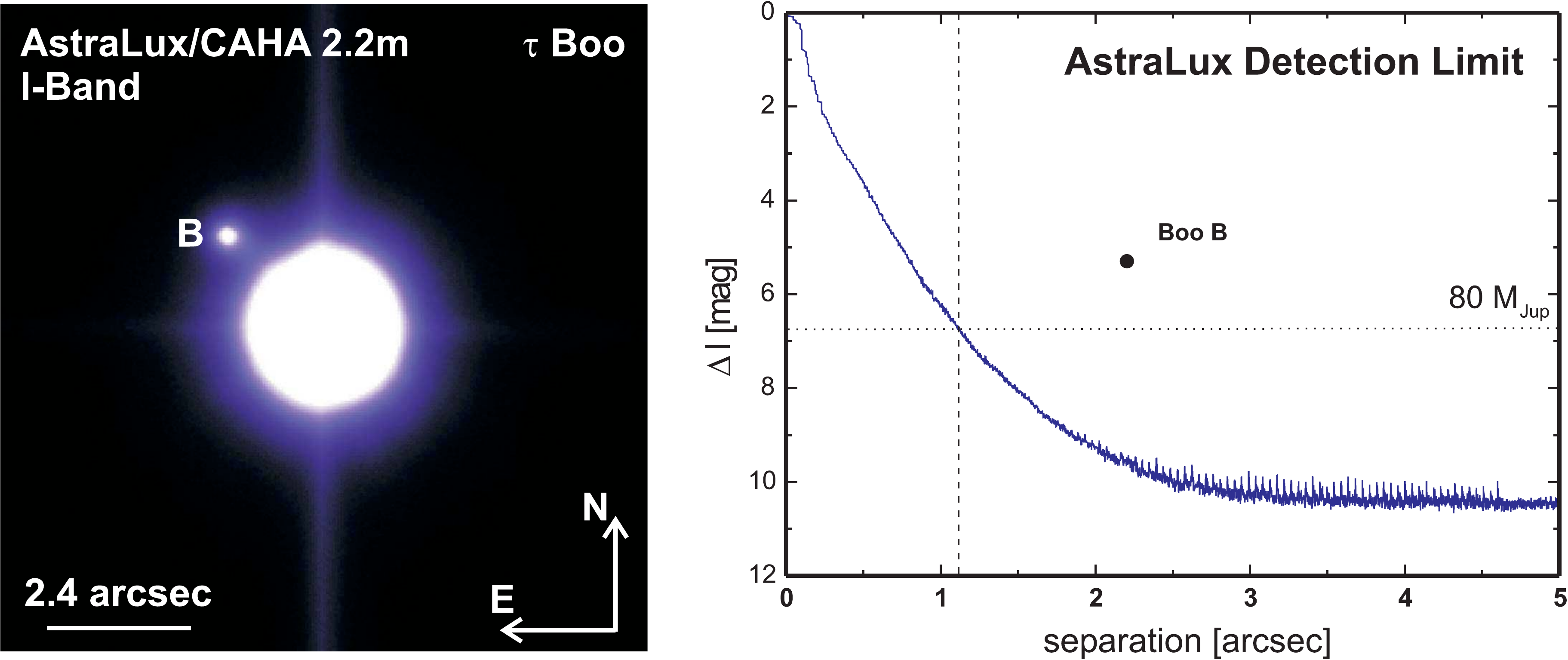}} \caption{As an example of our Lucky-Imaging
observations, we show here the exoplanet host star $\tau$\,Boo with its close stellar companion
$\tau$\,Boo\,B (0.44\,M$_\odot$), imaged by us with AstraLux. The companion is located about
2.2\,arcsec ($\sim$34\,AU of projected separation) north-east of its primary. The magnitude
difference between the companion and the planet host star is $\Delta I=5.3$\,mag. According to the
achieved AstraLux detection limit, companions like $\tau$\,Boo\,B can easily be imaged around all
our targets. In particular, beyond 1\,arcsec ($\sim$40\,AU of projected separation at the average
distance of our targets) all stellar companions ($>$80\,M$_{Jup}$) are detectable around all
targets.}
\end{figure}

\end{appendix}


\begin{thebibliography}{}
\bibitem[1]{1} Mugrauer, M., Seifahrt, A., Neuh\"{a}user, R., \& Mazeh, T.\ 2006, {\it MNRAS}, 373, L31
\bibitem[2]{2} Mugrauer, M., Seifahrt, A., Neuh\"{a}user, R., Mazeh, T., \& Schmidt, T.\ 2007, {\it IAUS}, 240, 638
\bibitem[3]{3} Mugrauer, M., Neuh\"{a}user, R., \& Mazeh, T.\ 2007, {\it A\&A}, 469, 755
\bibitem[4]{4} Mugrauer, M., Seifahrt, A., \& Neuh\"{a}user, R.\ 2007, {\it MNRAS}, 378, 1328
\bibitem[5]{5} Mugrauer, M., \& Neuh\"{a}user, R.\ 2009, {\it A\&A}, 494, 373
\end{thebibliography}
\end{document}